\documentclass[12pt]{article}
\usepackage{hyperref}
\usepackage[top=2cm,bottom=3cm,left=2cm,right=2cm]{geometry}
\usepackage{graphicx}
\usepackage{amsmath}
\usepackage{amssymb}

\numberwithin{equation}{section}
\numberwithin{figure}{section}

\def\eq#1{(\ref{eq:#1})}
\def\lineup{\!\!\!\!\!\!\!\! &&}

\newcommand{\Tr}{\mathop{\rm Tr}\nolimits}
\def\d{\partial}

\def\lineup{~&}

\def\UHP{\mathrm{UHP}}
\def\BCFT{\mathrm{BCFT}}
\def\tv{\mathrm{tv}}
\def\H{\mathcal{H}}

\def\M{{\bf M}}
\def\J{{\bf J}}
\def\p{{\bf p}}

\begin{document}

\begin{titlepage}

	\hfill \today
	\begin{center}
		\vskip 2cm
		
		{\Large \bf {Symplectic structure in open string field theory II:\\ Sliding lump}
		}
		
		\vskip 0.5cm
		
		\vskip 1.0cm
		{\large {Vin\'{\i}cius Bernardes$^{1}$, Theodore Erler$^{1}$, and Atakan Hilmi F{\i}rat$^{2}$ }}
		
		\vskip 0.5cm
		
		{\em  \hskip -.1truecm
			$^{1}$
			CEICO, FZU - Institute of Physics of the Czech Academy of Sciences \\
			No Slovance 2, 182 21, Prague 8, Czech Republic
			\\
			\vskip 0.5cm
			$^{2}$
			Center for Quantum Mathematics and Physics (QMAP),
			Department of Physics \& Astronomy, \\
			University of California, Davis, CA 95616, USA
			\\
			\vskip 0.5cm
			\tt \href{mailto:viniciusbernsilva@gmail.com}{viniciusbernsilva@gmail.com},
		\href{mailto:tchovi@gmail.com}{tchovi@gmail.com},
		 \href{mailto:ahfirat@ucdavis.edu}{ahfirat@ucdavis.edu} \vskip 5pt }
		
		\vskip 2.0cm
		{\bf Abstract}
		
	\end{center}
	\vskip 0.25cm
	\noindent
	\begin{narrower}
		\baselineskip15pt
		
		\noindent We use a new formula for symplectic structure to compute the momentum of an analytic lump solution moving at constant velocity in Witten's open string field theory. The computation gives a new way to determine the D-brane tension in string field theory. Using homotopy algebra technology, we prove that this tension must agree with the value implied by the on-shell action. 
		
	\end{narrower}
\end{titlepage}

\tableofcontents
\baselineskip15pt

\section{Introduction}

This is the second of three papers \cite{Bernardes1,Bernardes2} whose purpose is to develop a new formula for phase space symplectic structure \cite{Bernardes3} in open string field theory (open SFT). In the previous paper we studied rolling tachyon solutions in Siegel gauge \cite{Sen}, revisiting the calculation of Cho, Mazel, and Yin \cite{Cho}. That solution was perturbative and our description of it was mainly numerical. In this paper we focus our attention on a nonperturbative and analytic solution \cite{Schnabl,Erler}. 

We consider a D1-brane in bosonic string theory and let the tachyon condense to form a solitonic lump. The lump represents a D0-brane as seen from the degrees of freedom of the original D1-brane. Performing a Lorentz boost we obtain a ``sliding lump'' representing a D0-brane moving at constant velocity $v$. By shifting the origin, we can arrange it so that the D0-brane passes through the position $x_0$ at $t=0$. In this way we have a two dimensional phase space of time-dependent solutions parameterized by $(x_0,v)$. This phase space carries a nonzero symplectic structure which can be arranged in the form
\begin{equation}\Omega  = \delta p(v) \delta x_0, \end{equation}
where $p(v)$ is the spatial momentum of the sliding lump as a function of the velocity $v$. The spatial momentum should be given by 
\begin{equation}p(v) = \frac{m v}{\sqrt{1-v^2}},\end{equation}
where $m$ is the mass of the D0-brane. To goal of this paper is to confirm this by evaluating the symplectic structure using the analytic lump solution of \cite{Erler3,Erler4}. The interesting part is finding the lump mass. The mass is computed from the expression 
\begin{equation}m = \omega\big(ip_1\Phi,[Q_\Phi,\sigma]i J^{01}\Phi\big),\end{equation}
where $p_1,J^{01}$ are the translation and boost generators on the D1-brane worldvolume. This is a new way to determine D-brane tension which is not related in an obvious way to the action or the Ellwood invariant \cite{Ellwood,Kudrna}. Its analytic evaluation is nontrivial and exactly reproduces the expected mass of the D0-brane. 

In the final part of the paper we consider the symplectic structure of a boosted solution in an arbitrary $L_\infty$ field theory with Lorentz symmetry. We study the resulting mass formula using coalgebra techniques, giving a formal proof of equivalence to the mass obtained from evaluating the on-shell action. The proof requires taking some care of boundary contributions and gives an illustrative example of how Hamiltonian observables may be related in the language of $L_\infty$ algebras.

\subsubsection*{Conventions}

We use mostly plus metric and $\alpha'=1$. The ghost correlator is normalized as 
\begin{equation}
\langle c(z_1) c(z_2) c(z_3)\rangle_\UHP^\mathrm{bc} = z_{12}z_{13}z_{23},\ \ \ z_{ij}=z_i-z_j,
\end{equation}
and we use the left handed convention for the open string star product \cite{Erler2}. Commutators are graded with respect to Grassmann parity or, in section \ref{sec:general}, with respect to the even/odd parity~on~$\H$. 

\section{Sliding Lump}
\label{sec:sliding}

We consider a D1-brane in bosonic string theory. The worldsheet boundary conformal field theory (BCFT) takes the form
\begin{equation}
\BCFT = \BCFT_{X^0,X^1}\otimes \BCFT_\perp \otimes\BCFT_{bc}.
\end{equation}
The first factor represents the worldvolume of the D1-brane, and consists of noncompact timelike and spacelike free bosons $X^0(z,\overline{z})$, $X^1(z,\overline{z})$ subject to Neumann boundary conditions. This is the part of the BCFT where the tachyon condenses to form the lump.  The position and momentum zero mode operators are denoted by $x^\mu,p_\mu$ with $\mu=0,1$ and satisfy
\begin{equation}[x^\mu,p_\nu]= i\delta^\mu_\nu.\end{equation}
Sometimes space and time coordinates will be written without Lorentz indices as 
\begin{equation}x= x^1,\ \ \ t = x^0.\end{equation}
The second factor of the worldsheet theory is the ``transverse" BCFT, which we will imagine consists of 24 noncompact spacelike free bosons subject to Dirichlet boundary conditions. The exact form of the transverse BCFT is not very important because the formation of the lump does not excite transverse primary operators. What matters is that the central charge is $24$ and that the disk partition function is not zero 
\begin{equation}
\langle 1\rangle_\mathrm{disk}^\perp = Z_\perp\neq 0.
\end{equation}
The final factor of the worldsheet theory is the $bc$ ghost system of central charge~$-26$. The total BCFT has $c=0$. The dynamical string field  $\Psi$ is a Grassmann odd state of the BCFT at ghost number $1$.  We consider Witten's open bosonic string field theory~\cite{Witten}, with the action
\begin{equation}
S = - \frac{1}{g^2}\Tr\left(\frac{1}{2}\Psi Q\Psi+\frac{1}{3}\Psi^3\right),
\end{equation}
where $Q$ is the BRST operator, multiplication is carried out with Witten's open string star product, and the trace is defined by
\begin{equation}\Tr(A) = \langle I,A\rangle,\end{equation}
where $\langle,\rangle$ is the BPZ inner product and $I$ is the identity string field. The notation and conventions follow \cite{Erler}.  If the string field vanishes at infinity, the trace is BRST invariant and cyclic,
\begin{equation}\Tr(Q A) = 0,\ \ \ \ \Tr(AB) = (-1)^{AB}\Tr(BA).\end{equation}
Usually these properties are taken to hold without question, but in the context of the covariant phase space formalism we need to be careful about boundary terms \cite{Stettinger2,Firat2,Maccaferri2,Maccaferri3}. 

We want a lump solution traveling at velocity $v$. This can be found by performing a Lorentz boost of the stationary lump. To determine the correct boost we need to review some things from special relativity. Let $\Lambda$ denote a Lorentz boost by a velocity~$v$. It acts on the coordinates $(x,t)$ of the D1-brane as
\begin{subequations}
\begin{align}
\Lambda \circ x & = \gamma(x- vt),\label{eq:Lambda_x}\\
\Lambda \circ t & = \gamma(-vx+t),
\end{align}
\end{subequations}
where 
\begin{equation}\gamma = \frac{1}{\sqrt{1-v^2}}\end{equation}
is the Lorentz factor (the speed of light is one). Let $T(x,t)$ be the tachyon field on the D1-brane. The boost by $\Lambda$ will be
\begin{equation}
\Lambda\circ T(x,t) = T\big(\gamma(x- vt),\gamma(t- vx)\big).\label{eq:LambdaT}
\end{equation}
If $T(x,t)$ represents a tachyon lump whose center is located at $x=0$, the boost $\Lambda\circ T(x,t)$ will represent a lump whose center is localized at $x=vt$. Therefore the lump is now moving at a velocity $v$. It is useful to express the boost in terms of the rapidity, 
\begin{equation}v = \tanh \theta, \end{equation}
as
\begin{subequations}
\begin{align}
\Lambda \circ x & = (\cosh \theta) x- (\sinh \theta) t,\\
\Lambda \circ t & = -(\sinh\theta) x + (\cosh \theta) t.
\end{align}
\end{subequations}
Note that 
\begin{equation}
\left(\frac{\d}{\d\theta} + x\frac{\d}{\d t}+ t\frac{\d}{\d x}\right)\Lambda\circ x = \left(\frac{\d}{\d\theta} + x\frac{\d}{\d t}+ t\frac{\d}{\d x}\right)\Lambda\circ t = 0,
\end{equation}
which means that the boosted tachyon field satisfies
\begin{equation}
\frac{\d}{\d\theta} \Lambda\circ T(x,t) = -\left(x\frac{\d}{\d t}+ t\frac{\d}{\d x}\right)\Lambda\circ T(x,t).\label{eq:no_gen}
\end{equation}
We introduce the boost generator 
\begin{equation}
J^{01} = x^0 p^1 - x^1 p^0 = -i \left( t \frac{\d}{\d x}+x \frac{\d}{\d t}\right),
\end{equation}
so that \eq{no_gen} is written as
\begin{equation}
\frac{\d}{\d\theta} \Lambda\circ T(x,t) = -i J^{01} \big(\Lambda\circ T(x,t)\big),
\end{equation}
which implies
\begin{equation}
\Lambda \circ T(x,t) = e^{-i\theta J^{01}} T(x,t).
\end{equation}
This tells us how the sliding lump is created by applying the boost generator to the stationary lump. 

To include the massive states we need the full boost generator
\begin{equation}
J^{\mu\nu} = x^{[\mu}p^{\nu]} +\frac{i}{2}\sum_{n\in \mathbb{Z}-\{0\}} \frac{\alpha_n^{[\mu}\alpha_{-n}^{\nu]}}{n}.
\end{equation}
The bracket indicates antisymmetrization of indices (without $1/2!$). Any covariant operator $\mathcal{O}^\mu$ with a vector index satisfies
\begin{equation}[J^{\mu\nu},\mathcal{O}^\lambda] = -i \mathcal{O}^{[\mu}\eta^{\nu]\lambda}.\label{eq:Jcom}\end{equation}
The boost generator is a conserved charge on the worldsheet and can be expressed as a line integral of a conserved current. However the current is not holomorphic, so this works somewhat differently than usual. The boost generator is expressed 
\begin{equation}
J^{\mu\nu} = i\int_C \left(\frac{d\xi}{2\pi i} X^{[\mu}(\xi,\overline{\xi})\d X^{\nu]}(\xi)+\frac{d\overline{\xi}}{2\pi \overline{i}} X^{[\mu}(\xi,\overline{\xi})\overline{\d} X^{\nu]}(\overline{\xi})\right).
\end{equation}
The integration is performed over a contour $C$ in the upper half plane, and we do not use the doubling trick. The upper half plane is considered as a pair of real numbers $(a,b)$ with $b>0$. Complex coordinates are defined as $\xi = a+ib$ and $\overline{\xi} = a+\overline{i}b$, and in the end we equate $\overline{i} = -i$. The contour $C$ is oriented connecting two points on the real axis $a_1,a_2$ with $a_1>a_2$. Though the integrand is not holomorphic, the boost generator is independent of the choice of contour $C$. This follows from Stokes' theorem,
\begin{equation}
\int_D dz d\overline{z} \big(\overline{\d}\omega-\d\overline{\omega}\big) = \int_{\d D}\big(dz\,\omega + d\overline{z}\,\overline{\omega}\big),
\end{equation}
because the boost current is conserved:
\begin{equation}
\d\left(\frac{1}{2\pi \overline{i}} X^{[\mu}(\xi,\overline{\xi})\overline{\d} X^{\nu]}(\overline{\xi})\right) - \overline{\d}\left(\frac{1}{2\pi i} X^{[\mu}(\xi,\overline{\xi})\d X^{\nu]}(\xi)\right)=0.
\end{equation}
The boost generator must also be independent of where $C$ intersects the real axis. This requires that the boost current does not flow through the real axis:
\begin{equation}
\int da \left(\frac{1}{2\pi i} X^{[\mu}(a,a)\d X^{\nu]}(a)+ \frac{1}{2\pi \overline{i}} X^{[\mu}(a,a)\overline{\d} X^{\nu]}(a)\right) = 0.
\end{equation}
This holds because of the Neumann boundary condition on the D1-brane
\begin{equation}\d X^\mu(a) = \overline{\d}X^\mu(a),
\ \ \ \ a\in\mathbb{R}.\end{equation}
On the D0-brane this equation comes with a sign in the spatial direction, and the boost generator is not conserved. This expresses the fact that the boost generator modifies the boundary condition by making the D0-brane move. The momentum zero mode can also be written without the doubling trick
\begin{equation}
p^\mu = i\int_C\left(\frac{d\xi}{2\pi i} \d X^\mu(\xi) + \frac{d\overline{\xi}}{2\pi \overline{i}}\overline{\d}X^\mu(\overline{\xi})\right),
\end{equation}
and is conserved on the D1-brane by a similar argument as above. On the D0-brane, the spatial momentum is not conserved. 

Let $\Psi$ be a stationary lump solution describing a D0-brane sitting at the origin of the spatial coordinate $x$. The sliding lump solution is
\begin{equation}\Psi(x_0,v) = e^{-ix_0 p_1} e^{-i\theta J^{01}}\Psi.\label{eq:sliding_lump}\end{equation}
First we apply the boost generator to make the D0-brane move with velocity $v$. The D0-brane will then pass through the point $(x,t)=(0,0)$, as can be seen from \eq{LambdaT}. Second we apply the translation generator so that the lump passes through $x=x_0$ at $t=0$. Therefore we have a two dimensional phase space parameterized by the initial position $x_0$ and the initial velocity~$v$, as expected for a classical particle. We consider an explicit choice of stationary lump solution $\Psi$ later. 

\section{Symplectic structure}
\label{sec:symp}

The symplectic structure of Witten's open SFT is \cite{Bernardes1}
\begin{equation}
\Omega = -\frac{1}{g^2}\Tr\left(\frac{1}{2}\delta\Psi[Q,\sigma]\delta\Psi +\Psi[\sigma\delta\Psi,\delta\Psi]\right),\label{eq:Omega}
\end{equation}
where $\delta$ is the exterior derivative on solution space and $\sigma$ is a Grassmann even, ghost number zero operator called the {\it sigmoid}. The sigmoid is required to satisfy boundary conditions in the infinite past and infinite future:
\begin{equation}\lim_{t\to-\infty}\sigma = 0,\ \ \ \ \lim_{t\to\infty}\sigma = 1 ,\label{eq:sigma_bc}\end{equation}
where the limit refers to the region of support of the string field on which the sigmoid is acting. If the boundary conditions are obeyed, the symplectic structure formally does not depend on the choice of sigmoid. However, the sigmoid does effect how the symplectic structure is calculated. For analytic calculations it is advisable to adapt to the geometry of the Witten vertex as far as possible. In this context, the simplest choice of sigmoid is the one closest in spirit to Witten's original idea~\cite{Witten2} and is related to the canonical formulation of the theory in midpoint-lightcone time \cite{Bernardes1,Maeno,Erler5}. Here the sigmoid is given by a superposition of light-like plane wave vertex operators,
\begin{equation}\sigma = \int \frac{dE}{2\pi}\sigma(E) e^{iE X^+(i,\overline{i})},\end{equation}
inserted at the open string midpoint. The boundary conditions \eq{sigma_bc} will hold if the Fourier modes satisfy
\begin{equation}\lim_{E\to 0}\dot{\sigma}(E)=1,\end{equation}
where the dot indicates the time derivative, which in momentum space amounts to multiplication by $iE$. Because the Witten vertex is local in the lightcone coordinate of the midpoint, the sigmoid commutes through the open string star product. This implies that the interacting part of the symplectic structure vanishes because
\begin{equation}
[\sigma\delta\Psi,\Psi]=\sigma[\delta\Psi,\delta\Psi]=0.
\end{equation}
The quadratic part of the symplectic structure is defined by the midpoint insertion 
\begin{equation}
Q\sigma = [Q,\sigma ] = \gamma^+(i,\overline{i})\dot{\sigma},
\end{equation}
where we introduce
\begin{subequations}
\begin{align}
\gamma^\mu(z,\overline{z}) & = c\d X^\mu(z) + \overline{c}\overline{\d}X^\mu(\overline{z}),\label{eq:gammamu}\\
\dot{\sigma} & = \int \frac{dE}{2\pi}\dot{\sigma}(E)e^{iE X^+(i,\overline{i})}.
\end{align}
\end{subequations}
When the trace comes with a midpoint insertion, it is common to indicate it as a subscript,
\begin{equation}\Tr_\mathcal{O}(A) = \Tr\big(\mathcal{O}(i,\overline{i})A\big).\end{equation}
With this notation the symplectic structure \eq{Omega} simplifies to 
\begin{equation}
\Omega = -\frac{1}{2g^2}\Tr_{Q\sigma}\!\big(\delta\Psi^2\big).\label{eq:mplcOmega}
\end{equation}
The trace with midpoint insertion is BRST invariant and cyclic
\begin{equation}\Tr_\mathcal{O}(QA) = -(-1)^{\mathcal{O}}\Tr_{Q\mathcal{O}}(A),\ \ \ \ \Tr_\mathcal{O}(AB) = (-1)^{AB}\Tr_\mathcal{O}(BA),
\end{equation}
if boundary contributions can be ignored.

We now evaluate the symplectic structure for the sliding lump solution. The first step is to calculate the differential, which we write as 
\begin{equation}
\delta\Psi(x_0,v) = e^{-ix_0 p_1}e^{-i\theta J^{01}}\delta\Psi,
\end{equation}
where $\delta\Psi$ can be viewed as the differential in the rest frame of the lump. Plugging in \eq{sliding_lump} this is given by 
\begin{equation}
\delta\Psi = \Big(e^{i\theta J^{01}}e^{ix_0 p_1}\delta e^{-ix_0 p_1}e^{-i\theta J^{01}}\Big)\Psi.
\end{equation}
The exterior derivative acts on the phase space parameters to give
\begin{equation}
\delta\Psi =  \Big(-\delta x_0 \big(e^{i\theta J^{01}}ip_1 e^{-i\theta J^{01}}\big)-\delta\theta\, iJ^{01}\Big)\Psi.
\end{equation}
Raising an index, $p^1$ transforms under the boost as the spatial component of a vector. This can be read off from \eq{Lambda_x}, except we have to remember that we are boosting to velocity $-v$ to find the rest frame of the lump. Further lowering the Lorentz indices we infer
\begin{equation}e^{i\theta J^{01}}p_1 e^{-i\theta J^{01}} = \gamma(p_1- vp_0).\end{equation}
Since the stationary lump solution is time independent the contribution from $p_0$ drops out. We further express the differential of the rapidity in terms of the differential of the velocity using
\begin{equation}\delta v = \delta\tanh\theta = \mathrm{sech}^2\theta \delta\theta.\end{equation}
Since $\cosh\theta = \gamma$ we obtain
\begin{equation}\delta\theta = \gamma^2\delta v,\end{equation}
and
\begin{equation}
\delta\Psi = \Big(-\delta x_0 (\gamma i p_1)- \delta v (\gamma^2 i J^{01})\Big)\Psi.\label{eq:deltaPsi}
\end{equation}
The symplectic structure is
\begin{equation}
\Omega =  -\frac{1}{2g^2}\Tr_{Q\sigma}\!\big(\delta\Psi(x_0,v)^2\big)=  -\frac{1}{2g^2}\Tr_{Q\sigma}\!\Big[\big(e^{-ix_0 p_1}e^{-i\theta J^{01}}\delta\Psi)^2\Big].
\end{equation}
To get rid of the exponentials we transform everything backwards to the rest frame. This modifies the sigmoid, but since the symplectic form does not depend on the sigmoid we are free to ignore this. Therefore the symplectic structure of the sliding lump is simply \eq{mplcOmega} with $\delta\Psi$ given by~\eq{deltaPsi}. Plugging in gives\footnote{Since $Q\sigma$ is Grassmann odd, we must clarify that it appears as the leftmost insertion in the correlation function.}
\begin{equation}\Omega =  \frac{\gamma^3\delta v \delta x_0}{g^2} \Tr_{Q\sigma}\Big[(iJ^{01}\Psi)(ip_1\Psi)\Big].\end{equation}
Noting that
\begin{equation}\delta(\gamma v) = \gamma^3\delta v,\end{equation}
the symplectic structure can be expressed 
\begin{equation}\Omega = \delta p(v)\delta x_0, \end{equation}
where $p(v)$ is the relativistic momentum,
\begin{equation}p(v) = \frac{mv}{\sqrt{1-v^2}},\end{equation}
and the mass is given by
\begin{equation}m = \frac{1}{g^2}\Tr_{Q\sigma}\!\Big[(iJ^{01}\Psi)(ip_1\Psi)\Big].\label{eq:m}\end{equation}
In section \ref{sec:general} give a formal argument which relates this definition of mass to the value of the on-shell action. Presently however we will evaluate the formula directly for an analytic lump solution. The result should be the mass of the D0-brane, which is given as \cite{Sen2}
\begin{equation}m = \frac{Z_\mathrm{D0}}{\mathrm{vol}(X^0)}\frac{1}{2\pi^2g^2},\end{equation}
where $Z_\mathrm{D0}$ is the disk partition function of the D0-brane BCFT and $\mathrm{vol}(X^0)$ is the volume of the time coordinate.

\section{Lump mass}

To compute the mass we need an explicit lump solution. We consider the analytic lump solution given by the so-called intertwining construction of \cite{Erler3,Erler4}. This uses boundary condition changing operators, specifically Neumann-Dirichlet twist fields, to describe the shift in the background from the D1-brane to the D0-brane. Presently this is the only known analytic construction of a lump solution. 

\begin{figure}[t]
	\centering
	\includegraphics[scale=.7]{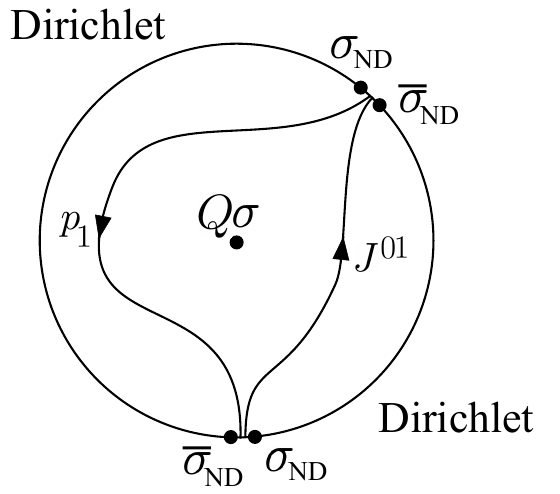}
	\caption{\label{fig:SympOSFT_II1} When evaluating the mass formula \eq{m} using the analytic lump solution, Neumann-Dirichlet twist fields $\sigma_\text{ND},\overline{\sigma}_\text{ND}$ collide and pinch the endpoints of the translation and boost generators, creating a logarithmic divergence. }
\end{figure}

Unfortunately, a direct evaluation of the mass formula \eq{m} appears to be indeterminate in this case. The problem comes from the Neumann-Dirichlet twist fields. The formula \eq{m} requires computing the overlap of the solution with itself, and in the process twist fields collide and pinch the endpoints of the $p_1$ and $J^{01}$ contours between them. This is illustrated in figure~\ref{fig:SympOSFT_II1}. The endpoints of the contours cannot be pulled apart because the boundary conditions are Dirichlet. The result is a logarithmic divergence created from twice integrating the double pole in the $\d X^1$-$\d X^1$ OPE near the boundary. Note that the divergence is also present for the flag state intertwining solution of~\cite{Erler4}, which otherwise deals with the collisions of boundary condition changing operators in a more robust fashion. For this reason we restrict our discussion on the simpler solution of \cite{Erler3}.

It is likely that this problem can be resolved by regularizing the collision of twist fields, but we do not pursue this.  Instead we use an alternative version of the mass formula where the logarithmic divergence is absent. To arrive at this version we assume that the sigmoid is constructed from a lightlike free boson of the form 
\begin{equation}X^+(z,\overline{z}) = \frac{1}{\sqrt{2}}\Big(X^0(z,\overline{z}) + X^2(z,\overline{z})\Big),\end{equation}
where $X^2(z,\overline{z})$ is a noncompact, spacelike free boson subject to Dirichlet boundary conditions which is taken from the transverse BCFT. We do not take the free boson from the worldvolume of the D1-brane because this would generate additional contractions between the sigmoid and the solution that complicate the calculation. An additional advantage is that the sigmoid has spatial translation symmetry
\begin{equation}[p_1,\sigma]=0,\label{eq:trans_sigmoid}\end{equation}
which will simplify the alternative mass formula below. 

We split the lump solution into two parts
\begin{equation}\Psi = \Psi_\tv +\psi,\end{equation}
where $\Psi_\tv$ is a tachyon vacuum solution on the D1-brane and $\psi$ creates the lump out of the tachyon vacuum. Far away from the D0-brane the lump solution should be very close to the tachyon vacuum, which means that $\psi$ will vanish.  We can assume that the tachyon vacuum is translation and boost invariant
\begin{equation}p_1\Psi_\tv  = J^{01}\Psi_\tv = 0.\end{equation}
The mass then reduces to 
\begin{equation}
m = \Tr_{Q\sigma}\!\Big[(iJ^{01}\psi)(ip_1\psi)\Big].\label{eq:m2}
\end{equation}
The alternative mass formula comes by considering 
\begin{equation}
\Tr_{Q\sigma}\!\Big[iJ^{01}\Big(\psi(ip_1\psi)\Big)\Big] = \Tr_{Q\sigma}\!\Big[(iJ^{01}\psi)(ip_1\psi)\Big]+\Tr_{Q\sigma}\!\Big[\psi\big(iJ^{01}ip_1\psi\big)\Big].
\end{equation}
We use the fact that $J^{01}$ is a derivation of the star product because it is a conserved charge. From \eq{Jcom} we have 
\begin{equation}[iJ^{01},ip_1] = -ip_0,\end{equation} 
which means that the second term can be replaced as
\begin{equation}
\Tr_{Q\sigma}\!\Big[iJ^{01}\Big((ip_1\psi)\psi\Big)\Big] =\Tr_{Q\sigma}\!\Big[(iJ^{01}\psi)(ip_1\psi)\Big]+ \Tr_{Q\sigma}\!\Big[\psi\big(ip_1 iJ^{01}\psi\big)\Big]-\Tr_{Q\sigma}\!\Big[\psi(ip_0\psi)\Big].
\end{equation}
The last term drops out because the lump solution is time independent. In the second term we pull out a total derivative $ip_1$ to write
\begin{equation}
\Tr_{Q\sigma}\!\Big[iJ^{01}\Big((ip_1\psi)\psi\Big)\Big] = \Tr_{Q\sigma}\!\Big[(iJ^{01}\psi)(ip_1\psi)\Big]+\Tr_{Q\sigma}\!\Big[ip_1\Big(\psi(iJ^{01}\psi)\Big)\Big]- \Tr_{Q\sigma}\!\Big[(ip_1 \psi)(iJ^{01}\psi)\Big].
\end{equation}
The first and last terms are equal to the mass \eq{m2} times $g^2$. Therefore
\begin{equation}
m = \frac{1}{2g^2}\bigg(\Tr_{Q\sigma}\!\Big[iJ^{01}\Big(\psi(ip_1\psi)\Big)\Big]- \Tr_{Q\sigma}\!\Big[ip_1\Big(\psi(iJ^{01}\psi)\Big)\Big]\bigg).
\end{equation}
Because $\psi$ vanishes at spatial infinity, and $Q\sigma$ vanishes at infinity in time, we can assume that the boost and translation generators annihilate the trace. Then means we can integrate by parts to apply these generators to the midpoint insertion. Since the sigmoid has translation symmetry we arrive at
\begin{equation}
m = \frac{1}{2g^2}\Tr_{iJ^{01}Q\sigma}\!\Big[(ip_1\psi)\psi\Big].\label{eq:m_alt}
\end{equation}
Now only the translation generator touches the open string boundary, and the mass formula has no logarithmic divergence.

The lump solution of \cite{Erler3} is given by
\begin{align}
\Psi_\tv & = \frac{1}{\sqrt{1+K}}c(1+K)Bc\frac{1}{\sqrt{1+K}},\\
\psi & = -\frac{1}{\sqrt{1+K}}c(1+K)\sigma \frac{1}{1+K}\overline{\sigma}(1+K)Bc\frac{1}{\sqrt{1+K}},
\end{align}
where $\Psi_\tv$ is the simple tachyon vacuum of \cite{Erler2}, $K,B$ and $c$ are string fields generating the $KBc$ subalgebra \cite{Erler,Erler2,Okawa}, and $\sigma$ and $\overline{\sigma}$ are string fields that shift from D1-brane to D0-brane boundary conditions and back. (These should not be confused with the sigmoid.) In correlation functions on the semi-infinite cylinder, $\sigma$ and $\overline{\sigma}$ represent insertions of Neumann-Dirichlet twist fields of weight $\frac{1}{16}$ in the $X^1$ BCFT,
\begin{equation}\sigma \, \to\, \sigma_\text{ND} e^{\frac{i}{4}X^0(0,0)},\ \ \ \ \overline{\sigma} \, \to\, \overline{\sigma}_\text{ND} e^{-\frac{i}{4}X^0(0,0)},
\end{equation}
together with plane wave vertex operators in the $X^0$ BCFT which make the total boundary condition changing operator a primary of weight zero. On account of this, the OPE of the boundary condition changing operators is regular and proportional to the identity operator. At the level of string fields, this reduces to the property 
\begin{equation}\overline{\sigma}\sigma = 1,\end{equation}
where $1$ is the identity string field on the D0-brane.

\begin{figure}[t]
	\centering
	\includegraphics[scale=.7]{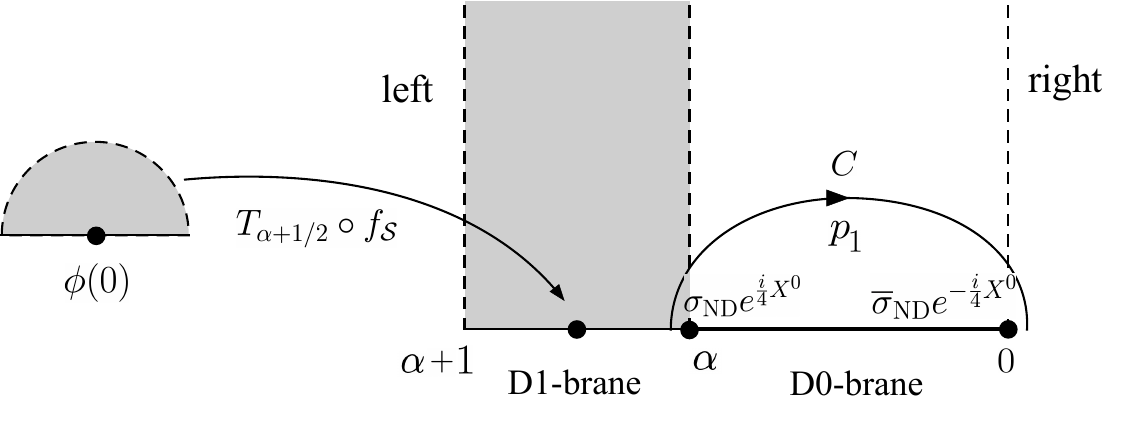}
	\caption{\label{fig:SympOSFT_II2} Illustration of the correlation function \eq{psOs}. The semi-infinite cylinder is displayed as a semi-infinite strip where the left and right edges are glued. The complex plane is shown so that the positive real axis increases to the left, which is natural in the left handed star product convention \cite{Erler2}. Note that different segments of the open string boundary carry D1-brane or D0-brane boundary conditions.}
\end{figure}

Plugging $\psi$ into \eq{m_alt} and making a few $KBc$ manipulations we arrive at the expression 
\begin{equation}
m=-\frac{1}{2g^2}\Tr_{iJ^{01}Q\sigma}\left[ip_1\left(\sigma\frac{1}{1+K}\overline{\sigma}\right)Bc\d c\, \sigma\frac{1}{1+K}\overline{\sigma}Bc\d c\right].\label{eq:m1st}
\end{equation}
To understand what the translation generator is doing we consider the simpler expression $ip_1\big(\sigma \Omega^\alpha\overline{\sigma}\big)$ where $\Omega^\alpha$ is a wedge state of width $\alpha$. Computing the overlap with a test state $\phi$ we find a correlation function on the semi-infinite cylinder
\begin{align}
& \big\langle \phi,ip_1\big(\sigma \Omega^\alpha\overline{\sigma}\big)\big\rangle\label{eq:psOs}\\
&\ \ = -\left\langle \big(T_{\alpha+\frac{1}{2}}\circ f_\mathcal{S}\circ \phi(0)\big)\int_C\left(\frac{dz}{2\pi i} \d X^1(z) + \frac{d\overline{z}}{2\pi \overline{i}}\overline{\d}X^1(\overline{z})\right)\big(\sigma_\text{ND} e^{\frac{i}{4}X^0(\alpha,\alpha)}\big)\big(\overline{\sigma}_\text{ND} e^{-\frac{i}{4}X^0(0,0)}\big)\right\rangle_{\!C_{\alpha+1}},\nonumber
\end{align}
where $C_L$ denotes the semi-infinite cylinder of circumference $L$, $f_\mathcal{S}(\xi)=\frac{2}{\pi}\tan^{-1}\xi$ is the local coordinate map defining the sliver frame, $T_a(z) = z+a$ is a translation on the cylinder by a constant $a$, finally, the translation generator is given by integrating the current over a contour $C$ connecting a point on the real axis a bit larger than $\alpha$ to a point just a bit smaller than $0$. An illustration of the correlation function is shown in figure \ref{fig:SympOSFT_II2}. The idea is to shrink the contour $C$ as tightly as possible around the twist fields. In the limit the contour can extend from the first twist field at $z=\alpha$ to the second twist field at $z=0$. The endpoints cannot be moved any closer because the open string boundary condition between the twist fields is Dirichlet. We can also collapse the contour onto the real axis, where the Dirichlet boundary condition allows us to equate
\begin{equation}\d X^1(a) = -\overline{\d} X^1(a),\ \ \ \ a\in \mathbb{R}.\end{equation}
 In this way we obtain
\begin{equation}
\int_C\left(\frac{dz}{2\pi i} \d X^1(z) + \frac{d\overline{z}}{2\pi \overline{i}}\overline{\d}X^1(\overline{z})\right)=-\frac{1}{i\pi}\int_0^\alpha da\, \d X^1(a),
\end{equation}
and
\begin{align}
& \big\langle \phi,ip_1\big(\sigma \Omega^\alpha\overline{\sigma}\big)\big\rangle \label{eq:p1sOs} \\ 
&\ \ \ \ \  = \frac{1}{i\pi}\left\langle \big(T_{\alpha+\frac{1}{2}}\circ f_\mathcal{S}\circ \phi(0)\big)\big(\sigma_\text{ND} e^{\frac{i}{4}X^0(\alpha,\alpha)}\big)\left(\int_0^\alpha da\, \d X^1(a)\right)\big(\overline{\sigma}_\text{ND} e^{-\frac{i}{4}X^0(0,0)}\big)\right\rangle_{C_{\alpha+1}}.\nonumber
\end{align}
The integration of $\d X^1$ over the Dirichlet segment represents the leading order deformation of the  the Dirichlet boundary condition from translation of the D0-brane. One thing to check is whether the collision between $\d X^1$ and the twist field leads to divergence. The OPE is
\begin{equation}
\d X^1(z)\sigma_\mathrm{ND}(0) = \frac{1}{z^{1/2}}\tau_\mathrm{ND}(0) + \text{less singular},
\end{equation}
where $\tau_\mathrm{ND}$ is the excited Neumann-Dirichlet twist field of weight $9/16$. The singularity in the OPE is integrable, so \eq{p1sOs} is not divergent. Noting that 
\begin{equation}\frac{1}{1+K} = \int_0^\infty d\alpha\, e^{-\alpha}\Omega^\alpha,\end{equation}
the conclusion is that the translation generator in \eq{m1st} acts as
\begin{equation}
ip_1\left(\sigma\frac{1}{1+K}\overline{\sigma}\right) = \frac{1}{i\pi} \sigma\frac{1}{1+K}\d X^1\frac{1}{1+K}\overline{\sigma},
\end{equation}
where $\d X^1$ on the right hand side is a string field representing a boundary insertion of $\d X^1(z)$ in correlation functions on the cylinder.

Plugging into \eq{m1st} the boundary condition changing operators cancel out, leaving
\begin{equation}
m=-\frac{1}{2\pi i g^2}\Tr_{iJ^{01}Q\sigma}\left[\frac{1}{1+K}\d X^1 \frac{1}{1+K} Bc\d c \frac{1}{1+K}Bc\d c\right].
\end{equation}
We have to keep in mind that $\sigma$ and $\overline{\sigma}$ have canceled in such a way as to leave D0-brane boundary conditions everywhere on the edge of the semi-infinite cylinder. To simplify the correlation function we want to eliminate the antighost insertions. Use
\begin{equation}Bc\d c B = [1+K,c]B\end{equation}
to write
\begin{equation}
m=\frac{1}{2\pi i g^2}\Tr_{iJ^{01}Q\sigma}\left[\frac{1}{1+K}B c\d X^1 \frac{1}{1+K} c\d c\right].
\end{equation}
Next we replace
\begin{equation}c\d c = Qc,\end{equation}
and integrate by parts to act the BRST operator on everything else. This is allowed because the midpoint insertion localizes the correlator to finite time and there is no spatial boundary because we are working on a D0-brane. Both the midpoint insertion and $c\d X^1$ are BRST invariant, so we only need $QB = K$ to arrive at
\begin{equation}
m=-\frac{1}{2\pi i g^2}\Tr_{iJ^{01}Q\sigma}\left[\frac{K}{1+K} c\d X^1 \frac{1}{1+K} c \right].
\end{equation}
Using
\begin{equation}\frac{K}{1+K} = 1-\frac{1}{1+K}\end{equation}
this becomes
\begin{equation}
m=\frac{1}{2\pi i g^2}\Tr_{iJ^{01}Q\sigma}\left[\frac{1}{1+K}c\d X^1 \frac{1}{1+K} c \right].
\end{equation}
The antighost insertions have now been eliminated.

Next we express this as a correlation function,
\begin{equation}
m = -\frac{1}{2\pi i g^2}\int_0^\infty d\alpha_1d\alpha_2 \, e^{-(\alpha_1+\alpha_2)}\Big\langle \big(T_{i\infty}\circ iJ^{01}Q\sigma \big)c(\alpha_2)c\d X^1(0)\Big\rangle_{C_{\alpha_1+\alpha_2}}^\mathrm{D0}.
\end{equation}
The semi-infinite cylinder of circumference $L$ is mapped to the upper half plane using
\begin{equation}g_L(z) = \tan \frac{\pi z}{L},\ \ \ \ \d g_L(z) = \frac{\pi}{L}\sec^2 \frac{\pi z}{L}.\end{equation}
This allows us to write
\begin{equation}
m = -\frac{1}{2\pi^2 i g^2}\int_0^\infty d\alpha_1 d\alpha_2 \, (\alpha_1+\alpha_2)\cos^2\left(\frac{\pi \alpha_2}{\alpha_1+\alpha_2}\right) e^{-(\alpha_1+\alpha_2)} \Big\langle\big(iJ^{01}Q\sigma\big) c(a) c\d X^1(0)\Big\rangle_\UHP^\mathrm{D0},\label{eq:m2nd}
\end{equation}
where the correlation function is now computed on the upper half plane with $c$ located at 
\begin{equation}
a = g_{\alpha_1+\alpha_2}(\alpha_2) = \tan\left(\frac{\pi \alpha_2}{\alpha_1+\alpha_2}\right).
\end{equation}
The action of the boost generator on $Q\sigma$ can be found from \eq{Jcom}. The result is
\begin{equation}
iJ^{01} Q\sigma = \frac{1}{\sqrt{2}}\Big(\gamma^1(i,\overline{i})\dot{\sigma} + X^1\gamma^+(i,\overline{i})\ddot{\sigma}\Big),
\end{equation}
where $\gamma^\mu$ is \eq{gammamu} and the double dot indicates the second derivative of the sigmoid with respect to midpoint lightcone time. The second derivative term does not contribute because $\d X^+$ finds nothing in the correlation function to form a nonzero contraction. Therefore
\begin{equation}
\Big\langle\big(iJ^{01}Q\sigma\big) c(a) c\d X^1(0)\Big\rangle_\UHP = \frac{1}{\sqrt{2}}\Big\langle\dot{\sigma}\big(c\d X^1(i) + \overline{c}\overline{\d} X^1(\overline{i})\big)c(a) c\d X^1(0)\Big\rangle_\UHP^\mathrm{D0}.
\end{equation}
The correlation function naturally factorizes into components from the $X^1$ BCFT, the ghost BCFT, and the $X^0$/transverse BCFTs:
\begin{align}
\Big\langle\big(iJ^{01}Q\sigma\big) c(a) c\d X^1(0)\Big\rangle_\UHP^\mathrm{D0} = \frac{1}{\sqrt{2}}& \Big(\big\langle \d X^1(i)\d X^1(0)\big\rangle_\mathbb{C}^{X^1}\big\langle c(i)c(a)c(0)\big\rangle_\mathbb{C}^{bc} \nonumber\\
& - \big\langle \d X^1(-i)\d X^1(0)\big\rangle_\mathbb{C}^{X^1}\big\langle c(-i)c(a)c(0)\big\rangle_\mathbb{C}^{bc}\Big)\big\langle\dot{\sigma}\big\rangle_\UHP^{X^0,\perp}.
\end{align}
We use the doubling trick to extend the $X^1$ and $bc$ correlators to the complex plane $\mathbb{C}$. Note the sign
\begin{equation}\overline{\d} X^1(\overline{i}) = -\d X^1(-i)\end{equation}
because of the Dirichlet boundary condition. Let 
\begin{equation} \langle 1\rangle_\UHP^{X^1}=Z_1,\ \ \ \ \text{(Dirichlet)}\end{equation}
be the disk partition function of the $X^1$ free boson BCFT when it is subject to Dirichlet boundary conditions. On the original D1-brane $X^1$ is subject to Neumann boundary conditions, and the disk partition function is usually equated with the volume of the $x^1$ coordinate. This is equivalent to normalizing the disk 1-point function~as
\begin{equation}
\big\langle e^{i kX^1(0,0)}\big\rangle_\UHP^{X^1} = 2\pi \delta(k),\ \ \ \ \text{(Neumann)}.
\end{equation}
This choice of normalization requires $Z_1 = 2\pi$ in order to obtain the correct ratio of tensions. Evaluating the $X^1$ and $bc$ correlators gives
\begin{align}
\Big\langle\big(iJ^{01}Q\sigma\big) c(a) c\d X^1(0)\Big\rangle_\UHP & = \frac{1}{\sqrt{2}}\bigg(\left(-\frac{1}{2}\frac{Z_1}{i^2}\right)(i-a)(i-0)(a-0) \nonumber\\
&\ \ \ \ \ \ \ \ \ \  -\left(-\frac{1}{2}\frac{Z_1}{(-i)^2}\right)(-i-a)(-i-0)(a-0)\bigg)\big\langle\dot{\sigma}\big\rangle_\UHP^{X^0,\perp}\nonumber\\
& = -\frac{i a^2 Z_1}{\sqrt{2}}\big\langle\dot{\sigma}\big\rangle_\UHP^{X^0,\perp}.
\end{align}
Finally we evaluate the sigmoid factor:
\begin{align}
\langle\dot{\sigma}\rangle_\UHP^{X^0,\perp}& = \int \frac{dE}{2\pi}\dot{\sigma}(E)\Big\langle e^{\frac{iE}{\sqrt{2}}X^0(i,\overline{i})}\Big\rangle_\UHP^{X^0}\Big\langle e^{\frac{iE}{\sqrt{2}}X^2(i,\overline{i})}\Big\rangle_\UHP^\perp.
\end{align}
This requires the 1-point functions
\begin{equation}
\big\langle e^{ik X^0(z,\overline{z})}\big\rangle_\UHP^{X^0} = 2\pi \delta(k),\ \ \ \ \big\langle e^{ik X^2(z,\overline{z})}\big\rangle_\UHP^\perp = Z_\perp (2\mathrm{Im}(z))^{k^2/4}.
\end{equation}
The normalization of the $X^0$ correlator is consistent with equating the disk partition function with the volume of time. In the transverse correlator $X^2$ carries Dirichlet boundary conditions, so there is no momentum conserving delta function. With this we can equate 
\begin{equation}Z_\mathrm{D0} = \mathrm{vol}(X^0) Z_1 Z_\perp.\end{equation} 
Plugging in the 1-point functions, integrating over $E$, and using $\dot{\sigma}(0)=1$ we obtain
\begin{equation}\langle\dot{\sigma}\rangle_\UHP^{X^0,\perp} = \sqrt{2}Z_\perp,\end{equation}
and 
\begin{equation}
\Big\langle\big(iJ^{01}Q\sigma\big) c(a) c\d X^1(0)\Big\rangle_\UHP = -ia^2 Z_1 Z_\perp.
\end{equation}
Finally, the mass \eq{m2nd} reduces to the integral 
\begin{equation}
m = \frac{Z_1 Z_\perp}{2\pi^2 g^2}\int_0^\infty d\alpha_1 d\alpha_2 \,(\alpha_1+\alpha_2)\sin^2\left(\frac{\pi \alpha_2}{\alpha_1+\alpha_2}\right)e^{-(\alpha_1+\alpha_2)}.
\end{equation}
Making a change of variables
\begin{equation}
L = \alpha_1+\alpha_2,\ \ \ \ \theta = \frac{\pi \alpha_2}{\alpha_1+\alpha_2},\ \ \ \ d\alpha_1d\alpha_2 = \frac{L}{\pi}dL d\theta,
\end{equation}
the integral becomes
\begin{equation}
m = \frac{Z_1 Z_\perp}{2\pi^3 g^2}\int_0^\infty dL \int_0^\pi d\theta L^2 \sin^2\theta \, e^{-L} .
\end{equation}
The integral of $\sin^2$ over a period gives the period divided by two. The integration over $L$ gives $\Gamma(3)=2$. In total
\begin{equation}
m = \frac{Z_1 Z_\perp}{2\pi^3 g^2}\cdot \frac{\pi}{2}\cdot 2 = \frac{Z_\mathrm{D0}}{\mathrm{vol}(X^0)}\frac{1}{2\pi^2 g^2}.
\end{equation}
This is precisely the expected mass of the D0-brane.

\section{General boosted solution}
\label{sec:general}

Another way to compute the mass formula \eq{m} is to transform it to the computation of the on-shell action. This can be done in Witten's SFT but the derivation is not especially intuitive. A more satisfying argument can be found by abstracting the problem and analyzing from the point of view of $L_\infty$ algebras.

We consider a generic Lagrangian field theory defined by a cyclic $L_\infty$ algebra \cite{Hohm,Jurco}. We will rely extensively on the coalgebra description of the theory. We use a version of the coalgebra formalism developed for applications to superstring field theory in \cite{Erler8,Erler6}. See~\cite{Erler6,Vosmera} for an introduction to the formalism. The dynamical field $\Phi$ is an element of a graded vector space $\H$. The vector space has an integer cohomological grading and an even/odd grading describing whether elements are commuting or anticommuting. The dynamical field $\Phi$ is grade 0 and commuting. The coalgebra formalism describes the theory in terms of operators acting on the tensor algebra of $\H$, denoted $T\H$. The action is 
expressed  \cite{Erler7}
\begin{equation}
S = -\int_0^1 ds\,\omega\!\left(\frac{\d \Phi(s)}{\d s},\pi_1\M \frac{1}{1-\Phi(s)}\right),
\end{equation} 
where $\Phi(s)\in\H$ is an element of $\H$ satisfying boundary conditions
\begin{equation}\Phi(0) = 0,\ \ \ \ \Phi(1) = \Phi.\end{equation}
Though it is not manifest, the action only depends on $\Phi(s)$ at $s=1$. This expression for the action invokes the following objects taken from the coalgebra formalism: 
\begin{itemize}
\item The group-like element of the tensor algebra generated by $\Phi(s)$:
\begin{equation}\frac{1}{1-\Phi(s)}= 1_{T\H} + \Phi(s) + \Phi(s)\otimes\Phi(s) + \cdots \end{equation}
\item The projection $\pi_n$ to the $\H^{\otimes n}$ component of the tensor algebra.
\item The Batalin-Vilkovisky (BV) inner product
\begin{equation}\omega(A,B) = \langle \omega|A\otimes B\end{equation}
This is a symplectic bilinear form on $\H$ of grade $-1$ which anticommutes.
\item The coderivation $\M$ representing the cyclic $L_\infty$ algebra of the theory. The products $M_n$ of the $L_\infty$ algebra are related to $\M$ through 
\begin{equation}M_n = \pi_1 \M \pi_n\end{equation}
The products carry grade $+1$ and anticommute. 
\end{itemize}
The coderivation $\M$ defines a cyclic $L_\infty$ algebra if 
\begin{subequations}
\begin{align}
\M^2\, \mathrm{Sym} & = 0,\ \ \ \ (L_\infty\text{ relations}),\\[.1cm]
\langle \omega|\pi_2\M & = 0,\ \ \ \ \text{(cyclicity)},\label{eq:cyclicity}
\end{align}
\end{subequations}
where $\mathrm{Sym}$ denotes the projection onto the symmetric component of $T\H$. The first identity encodes a hierarchy of quadratic relations between the products known as $L_\infty$ relations. The second identity is known as cyclicity, and is related to the statement that the integral of a total derivative vanishes. This is true provided that boundary contributions can be ignored. It is important to mention that the products are defined with an ``open string" normalization
\begin{equation}M_n = \frac{1}{n!}L_n,\end{equation}
where $L_n$ are the products appearing in \cite{Bernardes3}. This normalization allows us to discuss the coalgebra formalism for $L_\infty$ algebras in the same way as for $A_\infty$ algebras. The difference for $A_\infty$ algebras is that $\M$ is nilpotent even without projecting onto the symmetric part of the tensor algebra. The traditional description to $L_\infty$ algebras is based on the symmetrized tensor algebra \cite{Lada}, where the products are normalized as $L_n$.  This description is less natural for discussing cyclicity because the BV inner product is antisymmetric. 

The symplectic structure is \cite{Bernardes3}
\begin{equation}\Omega = \frac{1}{2}\omega\big(\delta\Phi,[Q_\Phi,\sigma]\delta\Phi\big),\end{equation}
where 
\begin{equation}
Q_\Phi A = \pi_1 \M\frac{1}{1-\Phi}\otimes A\otimes \frac{1}{1-\Phi},\label{eq:QPhi}
\end{equation}
is the kinetic operator for fluctuations around a solution $\Phi$. We want to evaluate the symplectic structure for a  boosted solution, which assumes that the theory has Poincar{\'e} symmetry. We assume that the symmetry is manifest, which means that translation and boost generators are realized as linear operators on $\H$
\begin{align}p_\mu:\H\to\H,\ \ \ \ J^{\mu\nu}:\H\to\H,\end{align}
which are grade 0, commuting, and satisfy the Poincar{\'e} algebra. Furthermore  
\begin{subequations}
\begin{align}
 [\p_\mu,\M] & = [\J^{\mu\nu},\M] \, = 0,\ \ \ \ \text{(symmetry)},\\ 
\langle\omega|\pi_2\p_\mu & = \langle \omega|\pi_2 \J^{\mu\nu} = 0,\ \ \ \ \text{(cyclicity)},
\end{align}
\end{subequations}
where $\p_\mu,\J^{\mu\nu}$ are the coderivations formed from $p_\mu,J^{\mu\nu}$. Cyclicity holds if boundary contributions can be ignored. Cyclicity for $\p_\mu$ is is exactly the statement that the integral of a total derivative is zero.

We consider a time-independent solution $\Phi$ and perform a boost and translation along time and space coordinates $(x^0,x^1)$. The boosted solution is
\begin{equation}
\Phi(x_0,v) = e^{-i x_0 p_1} e^{-i \theta J^{01}}\Phi.
\end{equation}
We can evaluate the symplectic structure analogously to section \ref{sec:symp}, giving
\begin{equation}
\Omega = \delta p(v)\delta x_0,
\end{equation}
where $p(v)$ is the relativistic momentum defined by the mass
\begin{equation}
m = \omega\big(ip_1\Phi,[Q_\Phi,\sigma]i J^{01}\Phi\big).
\end{equation}
Specializing to Witten's SFT with a midpoint-lightcone sigmoid, this reduces to \eq{m}.

We assume that $\Phi$ approaches a translation and boost invariant solution $\Phi_0$ towards spatial infinity
\begin{equation}p_0\Phi_0 = p_1\Phi_0=J^{01}\Phi_0 = 0.\end{equation}
We decompose
\begin{equation}\Phi = \Phi_0 + \varphi,\end{equation}
where $\varphi$ is a localized disturbance around the vacuum $\Phi_0$. The disturbance is time-independent
\begin{equation}p_0\varphi =0. \end{equation}
The mass is 
\begin{equation}
m = \omega\big(ip_1\varphi,[Q_{\Phi_0+\varphi},\sigma]i J^{01}\varphi\big).
\end{equation}
To go any further we need to use cyclicity. This requires taking care of boundary terms. We do not need to worry about spatial boundaries because $\varphi$ vanishes at spatial infinity. However we do need to worry about future and past boundaries. We deal with this using tau regularization~\cite{Bernardes3}. The tau regularization can be implemented in more than one way. We proceed by replacing all instances of $\varphi$ as 
\begin{equation}
\varphi \to \tau \varphi,
\end{equation}
where $\tau$ is a commuting, grade zero operator on $\H$ which vanishes in the infinite past and future but reduces to the identity at finite time. More precisely, this should be understood as the limit of a sequence of operators $\tau_n$ which satisfy
\begin{equation}
\lim_{n\to\infty}\tau_n = 1\ \ \text{(finite time)},\ \ \ \ \lim_{x^0\to\pm\infty} \tau_n = 0.
\end{equation}
We are free to replace $\varphi$ with $\tau\varphi$ because these expressions differ only in the infinite past and future, where the commutator with the sigmoid vanishes. However, when $\tau$ operates on $\varphi$ the fields really do vanish in the infinite past and future, and cyclicity should hold.  Using \eq{QPhi} and the assumption that the sigmoid is preserved through the BPZ inner product we arrive at
\begin{align}
m = \lineup  \omega\!\left(ip_1\tau\varphi,\pi_1\M \frac{1}{1-(\Phi_0+\tau\varphi)}\otimes \sigma iJ^{01}\tau\varphi \otimes \frac{1}{1-(\Phi_0+\tau\varphi)}\right) \nonumber\\
\lineup\ -\omega\!\left(\sigma ip_1\tau\varphi,\pi_1\M \frac{1}{1-(\Phi_0+\tau\varphi)}\otimes iJ^{01}\tau\varphi \otimes \frac{1}{1-(\Phi_0+\tau\varphi)}\right).
\end{align}
This is more convenient to write in terms of the $L_\infty$ structure expanded around the solution $\Phi_0$
\begin{equation}\M_{\Phi_0} = e^{-{\bf \Phi}_0}\M e^{{\bf \Phi}_0},\end{equation}
where ${\bf \Phi}_0$ is the coderivation derived from $\Phi_0$ viewed as a 0-product. The coderivation $\M_{\Phi_0}$ satisfies the relations of a cyclic $L_\infty$ algebra 
\begin{subequations}
\begin{align}
(\M_{\Phi_0})^2\, \mathrm{Sym} & = 0,\ \ \ \ (L_\infty\text{ relations}),\\[.1cm]
\langle \omega|\pi_2\M_{\Phi_0} & = 0,\ \ \ \ \text{(cyclicity)},
\end{align}
\end{subequations}
when boundary contributions can be ignored. Furthermore, because both $\M$ and $\Phi_0$ are translation and boost invariant we have
\begin{align}
 [\p_\mu,\M_{\Phi_0}] = [\J^{\mu\nu},\M_{\Phi_0}] = 0.
\end{align}
The group-like element can be simplified using 
\begin{equation}
\frac{1}{1-(\Phi_0+\tau\varphi)} = e^{{\bf \Phi}_0}\frac{1}{1-\tau\varphi},
\end{equation}
so that \eq{m2} is expressed
\begin{align}
m = \lineup  \omega\!\left(ip_1\tau\varphi,\pi_1\M_{\Phi_0} \frac{1}{1-\tau\varphi}\otimes \sigma iJ^{01}\tau\varphi \otimes \frac{1}{1-\tau\varphi}\right) \nonumber\\
\lineup\ -\omega\!\left(\sigma ip_1\tau\varphi,\pi_1\M_{\Phi_0} \frac{1}{1-\tau\varphi}\otimes iJ^{01}\tau\varphi \otimes \frac{1}{1-\tau\varphi}\right).
\end{align}
Using cyclicity of $\M_{\Phi_0}$ and antisymmetry of the BV inner product, we rewrite this so that the sigmoid appears only on the first entry:
\begin{align}
m = \lineup  \omega\!\left(\sigma i J^{01}\tau\varphi,\pi_1\M_{\Phi_0} \frac{1}{1-\tau\varphi}\otimes ip_1 \tau\varphi \otimes \frac{1}{1-\tau\varphi}\right) \nonumber\\
\lineup\ -\omega\!\left(\sigma ip_1\tau\varphi,\pi_1\M_{\Phi_0} \frac{1}{1-\tau\varphi}\otimes iJ^{01}\tau\varphi \otimes \frac{1}{1-\tau\varphi}\right).
\end{align}
Using
\begin{equation}
i\p_1\frac{1}{1-\tau\varphi} = \frac{1}{1-\tau\varphi}\otimes ip_1 \tau\varphi \otimes \frac{1}{1-\tau\varphi},\ \ \
 \ i\J^{01}\frac{1}{1-\tau\varphi} = \frac{1}{1-\tau\varphi}\otimes iJ^{01} \tau\varphi \otimes \frac{1}{1-\tau\varphi},
\end{equation}
this simplifies to 
\begin{align}
m = \lineup  \omega\!\left(\sigma iJ^{01}\tau\varphi,\pi_1\M_{\Phi_0} i\p_1 \frac{1}{1-\tau\varphi}\right) -\omega\!\left(\sigma ip_1\tau\varphi,\pi_1\M_{\Phi_0}i \J^{01} \frac{1}{1-\tau\varphi}\right).
\end{align}
Next commute $\p_1$ and $\J^{01}$ to the left past $\M_{\Phi_0}$. Further using cyclicity we arrive at
\begin{equation}
m = \omega\!\left(\Big[(i J^{01})\sigma (i p_1)-(ip_1)\sigma(i J^{01})\Big]\tau\varphi, \pi_1 \M_{\Phi_0}\frac{1}{1-\tau\varphi}\right).\label{eq:m3}
\end{equation}
We make an observation. Because $\varphi$ is a solution to the equations of motion expanded around $\Phi_0$, we have the identity 
\begin{equation}\M_{\Phi_0} \frac{1}{1-\varphi} = 0.\label{eq:varphiEOM}\end{equation}
This means that the mass formula as written in \eq{m3} is nonzero only because $\tau$ appears in front of~$\varphi$. Since $\tau$ differs from $1$ only in the infinite past and future, the formula \eq{m3} is purely a boundary term.

The operator in the first entry of the inner product can be written as
\begin{equation}
(i J^{01})\sigma (i p_1)-(ip_1)\sigma(i J^{01}) = [i J^{01},\sigma] i p_1 -[ip_1,\sigma] i J^{01} -\sigma [ip_1,i J^{01}].
\end{equation}
Because the first two terms have the sigmoid inside a commutator, their contribution is localized to finite time. For these terms we can equate $\tau=1$, which gives zero by the equations of motion~\eq{varphiEOM}. For the last term we use
\begin{equation}
[ip_1,i J^{01}] = i p_0
\end{equation}
to arrive at
\begin{equation}
m = -\omega\!\left(\sigma ip_0 \tau\varphi, \pi_1 \M_{\Phi_0}\frac{1}{1-\tau\varphi}\right).\label{eq:m4}
\end{equation}
This is nonzero not only because $\tau$ breaks the equations of motion, but because it breaks time translation symmetry of $\varphi$. 

The relation to the on-shell action appears after writing \eq{m4} in a form which is localized to finite time. This is achieved by a mechanism which is closely related to the reexpression of the closed string field theory action in a form which is proportional to the equations of motion \cite{Erler9}. The result is 
\begin{equation}m= i\int_0^1 ds\,\omega\!\left(\frac{\d\varphi(s)}{\d s},\pi_1\big(\p_0 \sigma \M_{\Phi_0}-\M_{\Phi_0}\sigma \p_0\big)\frac{1}{1-\varphi(s)}\right),\label{eq:Hamiltonian}\end{equation}
where $\varphi(s)$ satisfies boundary conditions
\begin{equation}\varphi(0) = 0,\ \ \ \ \varphi(1)=\varphi,\end{equation}
and $\sigma\M_{\Phi_0}$ and $\sigma \p_0$ are the coderivations derived by applying $\sigma$ to the output of the products in $\M_{\Phi_0}$ and $\p_0$. The right hand side of \eq{Hamiltonian} is actually the Hamiltonian functional \cite{Bernardes4}, where $\Phi_0$ is the state of zero energy. It is therefore natural that it should compute the mass of $\varphi$. The Hamiltonian is localized to finite time because $\sigma=0$ in the infinite past, while in the infinite future $\sigma =1$ and
\begin{equation}[\p_0,\M_{\Phi_0}]=0.\end{equation}
Therefore \eq{Hamiltonian} is defined independent of tau regularization. To relate to \eq{m4} we introduce the tau regularization through the replacement
\begin{equation}\varphi(s) \to \tau \varphi(s).\end{equation}
Then \eq{Hamiltonian} can be rewritten as
\begin{align}
m= -i\int_0^1 ds &\left[\omega\!\left( p_0 \frac{\d\tau \varphi(s)}{\d s},\sigma \pi_1\M_{\Phi_0}\frac{1}{1-\tau \varphi(s)}\right)\right.\nonumber\\
& \left.- \omega\!\left(\pi_1\M_{\Phi_0}\frac{1}{1-\tau\varphi(s)}\otimes \frac{\d \tau\varphi(s)}{\d s}\otimes \frac{1}{1-\tau\varphi(s)},\sigma p_0\tau\varphi(s)\right)\right],
\end{align}
where in the first term we used cyclicity of $\p_0$ and in the second term the cyclicity of $\M_{\Phi_0}$. Pull out the derivative in the second term and in the first term shift $\sigma$ to the first entry 
\begin{align}
m= -i\int_0^1 ds &\left[\omega\!\left(\sigma p_0 \frac{\d\tau \varphi(s)}{\d s},\pi_1\M_{\Phi_0}\frac{1}{1-\tau \varphi(s)}\right)- \omega\!\left(\frac{\d}{\d s}\pi_1\M_{\Phi_0}\frac{1}{1-\tau\varphi(s)},\sigma p_0\tau\varphi(s)\right)\right].
\end{align}
Antisymmetry of the BV inner product allows us to write this as the integral of a total derivative 
\begin{equation}
m= -i\int_0^1 ds \frac{\d}{\d s} \omega\!\left(\sigma p_0 \tau \varphi(s),\pi_1\M_{\Phi_0}\frac{1}{1-\tau \varphi(s)}\right).
\end{equation}
which leads to \eq{m4}. 

Finally we relate the Hamiltonian to the on-shell action. This requires commuting $p_0$ past the sigmoid and $\M_{\Phi_0}$ to act on $\varphi(s)$. Since the solution is time-independent we can choose $p_0\varphi(s)=0$. The mass simplifies to
\begin{equation}
m = \int_0^1 ds\, \omega\left(\frac{\d \varphi(s)}{\d s},\dot{\sigma}\pi_1\M_{\Phi_0}\frac{1}{1-\varphi(s)}\right),\label{eq:m5}
\end{equation}
where 
\begin{equation}\dot{\sigma} = [ip_0,\sigma].\end{equation}
The computation of \eq{m5} is unaffected by shifting the sigmoid in time, which means we can replace
\begin{equation}
\sigma\  \to\  e^{i t_0 p_0}\sigma e^{-i t_0 p_0}
\end{equation}
without changing the result. This is equivalent to replacing 
\begin{equation}\dot{\sigma} \ \to\ \frac{\d}{\d t_0}e^{i t_0 p_0}\sigma e^{-i t_0 p_0}\end{equation}
so that the mass is written
\begin{equation}
m = \frac{\d}{\d t_0}\int_0^1 ds\, \omega\left(\frac{\d \varphi(s)}{\d s},e^{i t_0 p_0}\sigma e^{-i t_0 p_0} \pi_1\M_{\Phi_0}\frac{1}{1-\varphi(s)}\right).
\end{equation}
Both sides of this equation are independent of $t_0$. Therefore we still obtain the mass if we integrate both sides over $t_0$ and divide by the volume of time,
\begin{equation}
m = \frac{1}{\mathrm{vol}(X^0)}\int_{-\infty}^\infty dt_0 \frac{\d}{\d t_0}\int_0^1 ds\, \omega\left(\frac{\d \varphi(s)}{\d s},e^{i t_0 p_0}\sigma e^{-i t_0 p_0} \pi_1\M_{\Phi_0}\frac{1}{1-\varphi(s)}\right).
\end{equation}
The integral of the total derivative gives boundary contributions
\begin{align}
m=\frac{1}{\mathrm{vol}(X^0)}\int_0^1 ds\, \omega\left(\frac{\d \varphi(s)}{\d s},\Big(\lim_{t_0\to\infty} e^{i t_0 p_0}\sigma e^{-i t_0 p_0}- \lim_{t_0\to-\infty}e^{i t_0 p_0}\sigma e^{-i t_0 p_0}\Big) \pi_1\M_{\Phi_0}\frac{1}{1-\varphi(s)}\right)\nonumber\\
\end{align}
The boundary conditions of the sigmoid imply that the first limit is 1 while the second limit is zero. Therefore 
\begin{equation}
m = \frac{1}{\mathrm{vol}(X^0)}\int_0^1 ds\, \omega\left(\frac{\d \varphi(s)}{\d s},\pi_1\M_{\Phi_0}\frac{1}{1-\varphi(s)}\right).
\end{equation}
which is the expected relation between the mass and the on-shell action expanded around $\Phi_0$.

\subsection*{Acknowledgments}

TE and AHF would like to thank D. Gross for hospitality at the KITP while carrying out part of this work. AHF thanks M. Rangamani for conversations. The work of AHF is supported by the U.S. Department of Energy, Office of Science, Office of High Energy Physics of U.S. Department of Energy under grant Contract Number DE-SC0009999, and the funds from the University of California. The work of VB and TE was supported by the European Structural and Investment Funds and the Czech Ministry of Education, Youth and Sports (project No. FORTE—CZ.02.01.01/00/22\_008/0004632). This research was supported in part by grant NSF PHY-2309135 and the Gordon and Betty Moore Foundation Grant No. 2919.02 to the Kavli Institute for Theoretical Physics~(KITP).

\end{document}